\documentclass[aps,prl,superscriptaddress,twocolumn,showpacs] {revtex4}

\usepackage{bm}
\usepackage{graphicx}
\usepackage{amsmath}
\usepackage{color}
\begin{document}

\title{Staggered magnetization and entanglement enhancement by
magnetic impurities in $S=1/2$ spin chain}
\author{Tony Apollaro}
\affiliation{Dipartimento di Fisica, Universit\`a di Firenze,
    Via G. Sansone 1, I-50019 Sesto Fiorentino (FI), Italy}
\affiliation{CNISM - Consorzio Nazionale Interuniversitario per le Scienze Fisiche della Materia}

\author{Alessandro Cuccoli}
\affiliation{Dipartimento di Fisica, Universit\`a di Firenze,
    Via G. Sansone 1, I-50019 Sesto Fiorentino (FI), Italy}
\affiliation{CNISM - Consorzio Nazionale Interuniversitario per le Scienze Fisiche della Materia}

\author{Andrea Fubini}
\affiliation{Dipartimento di Fisica, Universit\`a di Firenze,
    Via G. Sansone 1, I-50019 Sesto Fiorentino (FI), Italy}
\affiliation{CNISM - Consorzio Nazionale Interuniversitario per le Scienze Fisiche della Materia}

\author{Francesco Plastina}
\affiliation{Dip. Fisica, Universit\`a della Calabria, \& INFN -
Gruppo collegato di Cosenza, 87036 Arcavacata di Rende (CS) Italy}

\author{Paola Verrucchi}
\affiliation{ Centro di Ricerca e sviluppo {\it Statistical Mechanics and
Complexity} INFM-CNR}
\affiliation{Dipartimento di Fisica, Universit\`a di Firenze,
    Via G. Sansone 1, I-50019 Sesto Fiorentino (FI), Italy}

\date{\today}

\begin{abstract}

We study the effects of a magnetic impurity on the behavior of a
$S=1/2$ spin chain.
At $T=0$, both with and without an applied uniform magnetic field,
an oscillating magnetization appears, whose decay with the distance
from the impurity is ruled by a power law. As a consequence,
pairwise entanglement is either enhanced or quenched,
depending on the distance of the spin pair with respect to the
impurity and on the values of the magnetic field and the intensity
of the impurity itself. This leads us to suggest that acting on such 
control parameters, an adiabatic manipulation of the entanglement
distribution can be performed. The robustness of our
results against temperature is checked, and suggestions about
possible experimental applications are put forward.
\end{abstract}

\pacs{75.10.Pq, 75.10.Jm, 03.67.Bg, 75.30.Hx}

\maketitle

The prospect of quantum technology, i.e., the design and
realization of quantum devices, is triggering an increasing
interest in the manipulation of quantum systems in view of several
applications, amongst which is quantum information
processing~\cite{AmicoEtal08}. Despite the large variety of
experimental proposals, the relevant properties of most of such
devices can be adequately described by a few classes of models.
Quantum low-dimensional spin models constitute one such class:
They may describe different physical systems and, at the same
time, their Hamiltonian is simple enough to allow an easy
recognition of the parameters responsible for specific effects, so
that simple controls may be designed to drive focused
manipulations. Sticking to strictly one-dimensional models, a
possibility of manipulation is offered by the introduction of
local impurities, either magnetic or not, which break the
translational symmetry. In the last decade, several studies have
been devoted to low-dimensional systems with inhomogeneities~
\cite{EggertA9295,TakigawaEtal97,MartinsEtal97,FujiwaraEtal98,
TedoldiSH99,NishinoEtal00,AletS00,RommerE00,SirkerEtal07,BottiRST00,EggertEtal07},
revealing peculiar phenomena which occur in the vicinity of the
impurity.

In this paper, we consider the less investigated case of magnetic
impurities, which can couple both with neighboring spins and/or with
external magnetic fields, offering a remarkable control flexibility. In
particular, aimed at possible application in the realm of quantum
computation, we consider a single impurity in the $S=1/2$ XX chain
in a transverse field in its quasi-long-range ordered phase. In the
absence of impurity and due to the high symmetry of the exchange
interaction, such model displays values of bipartite entanglement
larger than those observed in less symmetric models~\cite{AmicoEtal06}.
The impurity is modelled in terms of an additional field located at one
precise site of the chain. We study the ground-state properties and
thermodynamic behavior when the field and/or the magnetic strength
of the impurity are varied.

At zero temperature, we find an infinite penetration depth for the
effects of the magnetic impurity, namely it produces spatial
modulations of the spin configuration which decay polynomially
with the distance from the impurity itself. A staggered
magnetization and an oscillating susceptibility result, as a most
general consequence of translational symmetry
breaking~\cite{EggertA9295}. One striking effect of the emergence
of such spin structure is the local enhancement of pairwise
entanglement: When no uniform field is applied, the
concurrence~\cite{Wootters98,AmicoEtal04} of the spin-pair next to
the impurity doubles with respect to the translational invariant
case and, consequently, an entanglement re-distribution takes
place all over the chain. With finite magnetic field,
there always exists a most entangled spin pair, which can be moved
along the chain by tuning either the field or the impurity
strength. These parameters can thus play the role of "knobs" for
the manipulation of both the magnetic and the entanglement
properties of the chain. Furthermore, as they appear in terms that
commute with the Hamiltonian, adiabaticity is guaranteed while
tuning their values. The essential features of the above picture,
and in particular the possibility of selectively enhance pairwise
entanglement, survive at finite temperature. This gives a
concrete possibility to experimentally test our findings.

Let us consider the Hamiltonian
\begin{equation}
\frac{\cal{H}}{J}={-}\sum_{n=-\frac{N}{2}}^{\frac{N}{2}-1}
\left[
\frac{1}{2} (\sigma^x_n\sigma^x_{n+1}{+}\sigma^y_n\sigma^y_{n+1})
{+}h\, \sigma^z_n \right]
{+}\alpha\,\sigma^z_0~,
\label{e.H}
\end{equation}
where $\sigma^\varepsilon_n$ ($\varepsilon=x,y,z$) are the the Pauli
operators for the spin at site $n$, and periodic boundary
conditions are assumed. A uniform magnetic field $H$ is applied
along the quantization axis, which takes values such that $h\equiv
g_{_{\rm L}} \mu_{_{\rm B}} H/J\in[-1,1]$; the exchange integral
$J$ is set to unity throughout the paper. The magnetic impurity is
located at site $0$ and represented by a local magnetic field
$\alpha$ \cite{note1}. As ${\cal H}$ can
describe several physical systems beyond the mere spins (e.g.
atoms loaded in optical lattices), one may think of
correspondingly several way of introducing a localized defect.
In general the impurity breaks the translational symmetry
and as $\alpha\rightarrow\infty$ it renders the
system equivalent to an open-end chain. 

The Hamiltonian (\ref{e.H}) can be diagonalized by standard
methods \cite{LiebSM61,ApollaroP06}, resulting in an energy band
and in a localized level
\begin{equation}
{\cal{H}}=\sum_k E_k g_k^\dag g_k+E_{_{\rm \lambda}} g_{_{\rm
\lambda}}^\dag g_{_{\rm \lambda}}~. \label{e.Hdiag}
\end{equation}
Here, $E_{_{\rm \lambda}}=h-(\text{sgn}\,\alpha)\sqrt{1+\alpha^2}$
is the energy of the localized state, while the corresponding
fermion operator is a combination of the Jordan-Wigner
ones \{$c_n^{\dag}$\}
\begin{equation*}
g_{\lambda}^{\dag}=\sum_{n}\varphi_{n,\lambda}c_n^{\dag}~;\qquad
 \varphi_{n,\lambda}=
-\frac{\sqrt{|\alpha|}\,(\text{sgn}\,\alpha)^n}{(1+\alpha^2)^{\frac{1}{4}}}
e^{-|n|/\xi} ~,
\end{equation*}
with $\xi^{-1}\equiv-\ln(\sqrt{1-\alpha^2}-|\alpha|)$. Concerning
the band, $E_k{=}h{-}\cos\theta_k$ is the unperturbed dispersion
relation ($\theta_k=\frac{2 \pi
k}{N},k=-\frac{N}{2},..,\frac{N}{2}-1$), while
\begin{equation}
g_k^{\dag}= \sum_{n}\frac{\varphi_{n,k}}{\sqrt{N}}
c_n^{\dag}~;\qquad \varphi_{n,k}= e^{i\theta_kn}-\frac{\alpha \,
e^{i |\theta_k|n}}{i\sin|\theta_k|+\alpha}~, \label{dacit}
\end{equation}
The impurity is seen to produce {\it i)} the appearance of a
discrete energy level (lying above or below the band depending on
the sign of $\alpha$), exponentially localized in space with a
characteristic length $\xi$;
and {\it ii)} the distortion of the energy-band states, whose
amplitudes, besides the plane wave term, contain a back-scattering
contribution due to the defect. It is precisely the interference
between these two terms that gives rise to the oscillating
patterns discussed below in both the magnetization and the
susceptibility.

Using the Wick's theorem, we calculate one- and two-points
correlation functions which, due to the lack of translational
symmetry, cannot be expressed as Toeplitz determinants, so that
their expressions become more and more involved with the relative
site distance. We find:
\begin{align}
&\langle{\sigma^z_r}\rangle{=}1{-}2\sum_{\eta\in{\cal F}}|\varphi_{r,\eta}|^2~,~
\langle{\sigma^x_r \sigma^x_{r+1}}\rangle{=}
2 Re\sum_{\eta\in{\cal F}} \varphi_{r,\eta}^*\varphi_{r+1,\eta}~,
\nonumber\\
&\langle{\sigma^z_r\sigma^z_s}\rangle
=1-2\sum_{\eta\in{\cal F}}
\left(|\varphi_{r,\eta}|^2+|\varphi_{s,\eta}|^2\right) +
\nonumber
\\
&\,\,\,\,\,\,\,+4\sum_{\{\eta,\nu\}\in{\cal F}, \eta\ne \nu}
\left(|\varphi_{s,\eta}|^2 |\varphi_{r,\nu}|^2 - \varphi_{r,\eta}
\varphi_{s,\eta}^*\varphi_{r,\nu}^*\varphi_{s,\nu}\right)
\label{e.Szz}
\end{align}
where ${\cal F}$ is the set of filled states. Rotational symmetry
gives $\langle{\sigma_r^x}\rangle=\langle{\sigma_r^y}\rangle=0$ and
$\langle{\sigma^x_r \sigma^x_{r+s}}\rangle=\langle{\sigma^y_r
\sigma^y_{r+s}}\rangle$; moreover, reflection symmetry with 
respect to site 0 allows us to consider only $r\geq0$ in the following.

\begin{figure}
\begin{minipage}[c]{1\linewidth}
\includegraphics[height=0.40\linewidth]{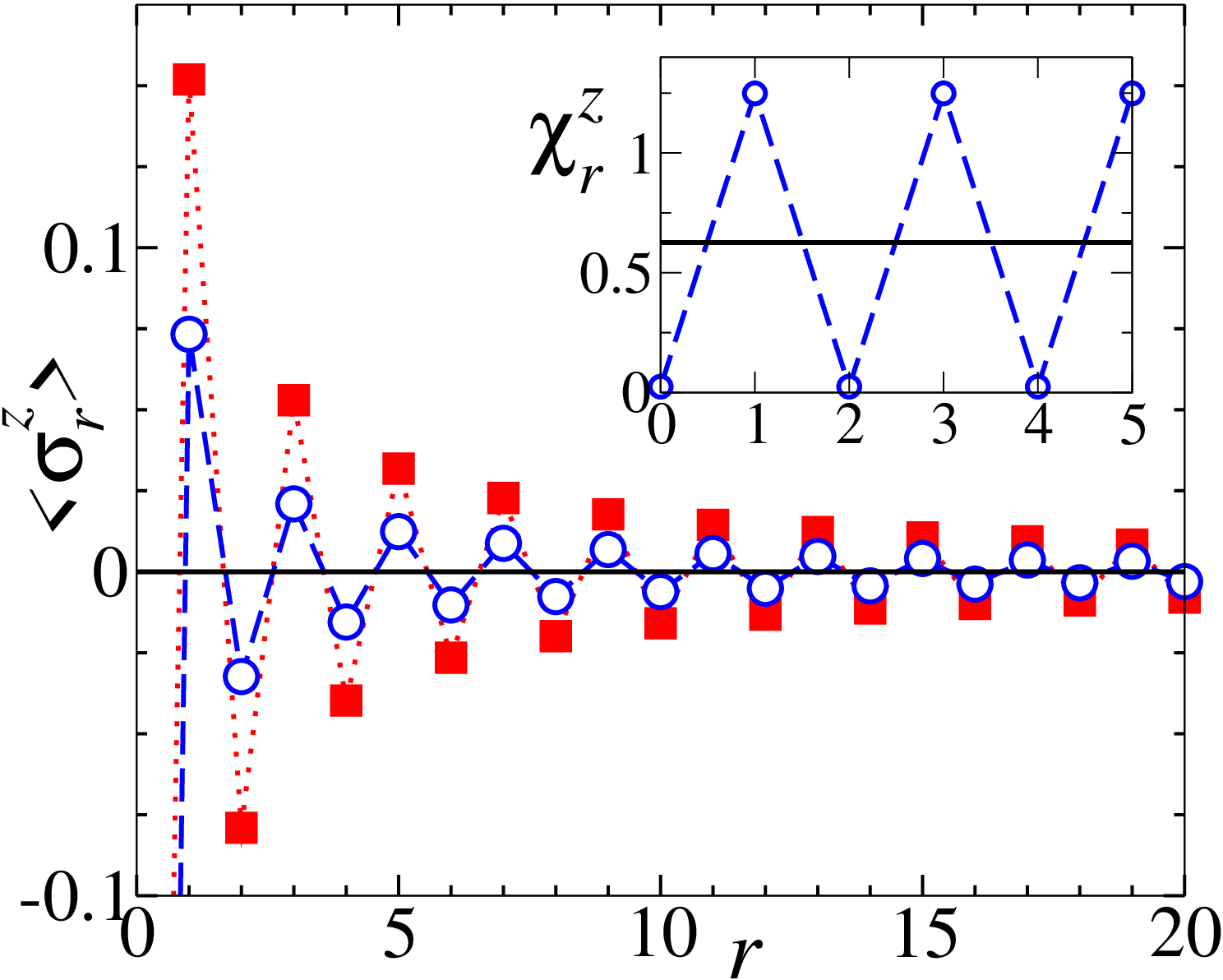}
\includegraphics[height=0.40\linewidth]{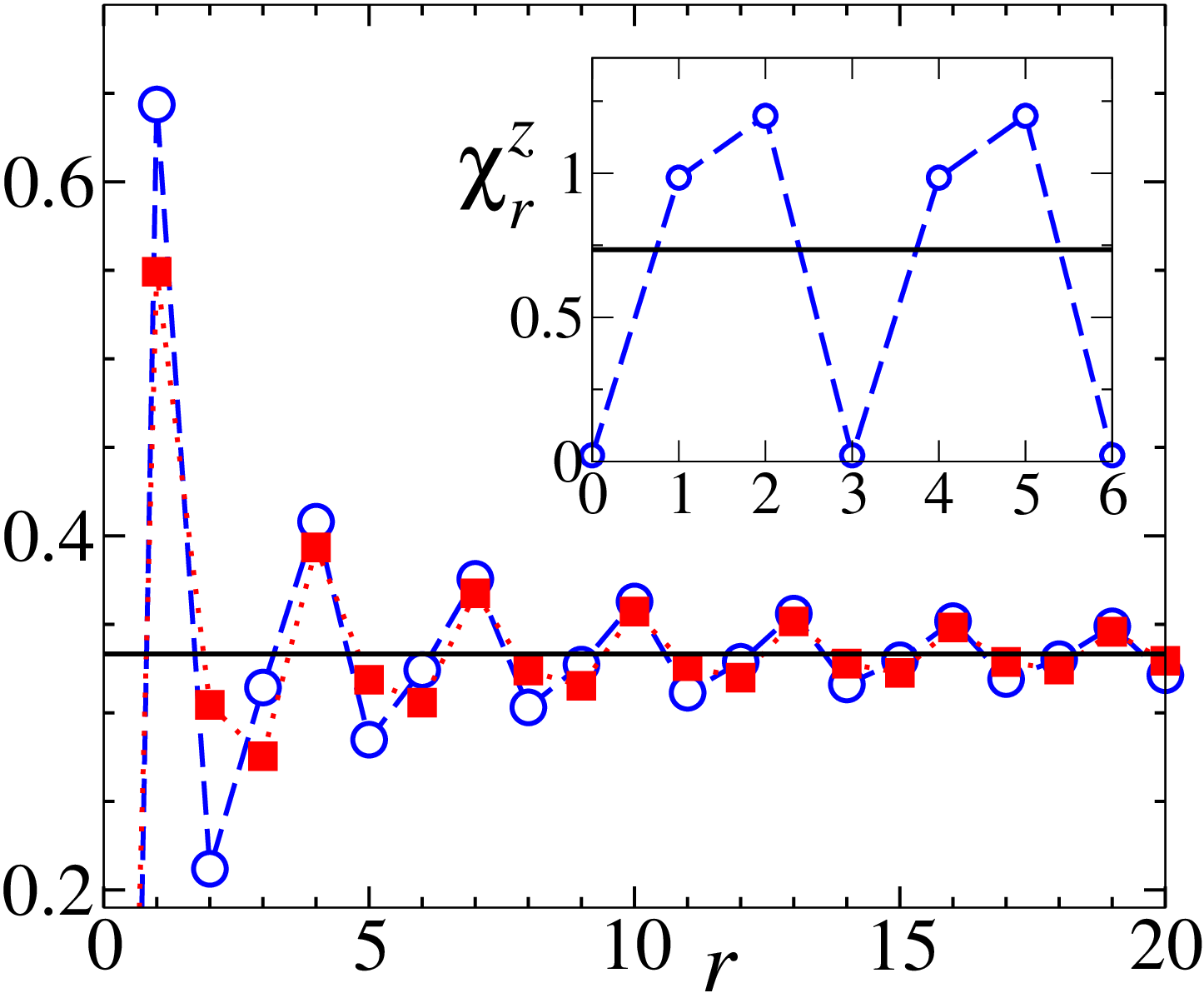}
\end{minipage}
\caption{$\langle{\sigma_r^z}\rangle$ at $h=0$ (left) and $h=0.5$ (right), for $\alpha=0$ (continuous line), $\alpha=1$ (filled square) and $\alpha=5$ (open circles). The insets show the local susceptibility for $\alpha=5$.} \label{f.sz}
\end{figure}

In the thermodynamic limit, the local magnetization at zero
temperature takes the form
\begin{equation}
\langle{\sigma_r^z}\rangle{=}
\langle{\sigma^z}\rangle{+}
\langle{\sigma_r^z}\rangle_{_\lambda}{+}
\frac{2\alpha}{\pi}\!\!\!\int_{0}^{\theta{_F}}
\!\!\!\!\!\!\!d\theta\frac{\alpha\cos(2r\theta)+\sin\theta\sin(2r\theta)}
      {\alpha^2+\sin^2\theta}~,
\label{e.Szthermo}
\end{equation}
with $\theta_{\rm F}\equiv\cos^{-1}h$ and where
$\langle{\sigma^z}\rangle=1-\frac{2\theta_{\rm F}}{\pi}$ gives the result
without defect, $\langle{\sigma_r^z}\rangle_\lambda=-2|\varphi_{r,\lambda}|^2$ is
due to the localized state (this term is present only for
$\alpha>0$), while the last term is due to the interference
mechanism mentioned above. The essential finding embodied in
Eq.~(\ref{e.Szthermo}) is the emergence of an oscillating
magnetization in the vicinity of the impurity, as shown in
Fig.~\ref{f.sz}.

Several works~\cite{EggertA9295,MartinsEtal97,TakigawaEtal97,%
FujiwaraEtal98, TedoldiSH99,%
NishinoEtal00,AletS00,RommerE00,SirkerEtal07,BottiRST00}
have evidenced that finite
staggered moments are induced by open ends in antiferromagnetic
spin chains
and that the spatial decay of the related magnetization is ruled
by the (bulk) correlation length of the infinite, translational
invariant, system. Previous analysis, however, referred to spin chains in disordered phases, with exponentially
decaying correlations,
either because of the finite temperature or due the symmetry of
the Hamiltonian itself, as in the case of the Haldane chains.
On the contrary, our model is independent of the sign of the
exchange coupling and, at $T=0$, it is in a quasi-long-range 
ordered phase with an infinite
correlation length and a much slower (power law) decay of the
correlation functions: The effects of the impurity are
consequently felt by more distant spins along the chain.
Indeed, there exists a precise analogy~\cite{RommerE00}
between our model and the case of a one-dimensional fermionic
system in the presence of one impurity, where Friedel oscillations
occur~\cite{Friedel58}: For $\alpha\gg1$ and
$r\gg1$ Eq.~(\ref{e.Szthermo}) gives the same $\frac{1}{r}$ behavior
predicted for the oscillations of the fermion density in a
Luttinger liquid with a defect~\cite{EggerG95}. It is noteworthy
that, at variance with the case of the XX chain with locally
inhomogeneous exchange interaction~\cite{RommerE00}, the presence
of the magnetic impurity gives rise to Friedel-like oscillations
even when no uniform field is present.

In general, for a fixed $h$, the value of $\alpha$ determines the
absolute value of the local magnetization, while, independently of
$\alpha$, the field fixes the spatial periodicity $p$ of the
oscillations (see Fig.~\ref{f.sz}). In fact, we see that
$p$ coincides with the period of the connected correlation
functions of the translational-invariant model, i.e.
$p=\frac{\pi}{\theta_{\rm F}}$ \cite{BarouchM71}. This behavior is
understood by observing that the local switching-on of
$\langle{\sigma^z_r}\rangle$ under the effect of the local field
at site $0$ is mediated by the generalized magnetic susceptibility,
which in turns has the same spatial dependence of the correlation
functions.

Let us now consider the local magnetic susceptibility. From
Eq.~(\ref{e.Szthermo}), we get
$\chi_r(\alpha,h)=2/(\pi\sqrt{1-h^2})+\chi^{\rm alt}_r(\alpha,h)$,
where the alternating term is
\begin{equation}
\chi^{\rm alt}_r(\alpha,h)=-2\alpha\frac
{\alpha \cos(2r\theta_{\rm F})+\sqrt{1-h^2}\sin(2r\theta_{\rm F})}
{\pi(1+\alpha^2-h^2)\sqrt{1-h^2}}~.
\label{e.altsusc}
\end{equation}
The superposition of a uniform term and a spatially oscillating
one, is fully analogous with the results relative to the
spin-$\frac{1}{2}$ Heisenberg antiferromagnetic chain with open
ends \cite{TakigawaEtal97} or bond impurities \cite{NishinoEtal00}.
The alternating term survives throughout the chain with a spatial
structure that, being $r$ an integer, follows peculiar, beating-like,
patterns, depending on the value of $h$. When $h=0$, one has
$\chi_r(\alpha,0)=
\frac{2}{\pi}\left(1-\frac{(-1)^r\alpha^2}{1+\alpha^2}\right)$
meaning that, for $\alpha\gg1$, odd-labelled spin increase their
susceptibility up to the value $\frac{4}{\pi}$, while even-ones
have zero susceptibility.

Noticeably, $\chi_r(\alpha,h)$  does not diverges at the critical
point $h=1$. In fact, while in the translational invariant model the
susceptibility diverges as $(1-h)^{-1/2}$ for $h\to 1$, we find
here that in the same limit $\chi_r(\alpha,h) \simeq \frac{2
\sqrt{2(1-h)}}{\pi}\left(\frac{1+2r\alpha}{\alpha^2+2(1-h)}\right)$:
As a consequence of the translational symmetry breaking,
singularity is cancelled and all susceptibilities vanish at the
critical point.

The spatial modulation of the magnetic susceptibility described in
Eq.~(\ref{e.altsusc}), offers a chance to selectively act on local
spins. This possibility is at the hearth of the following
discussion about local enhancement and possible transfer of
entanglement.

\begin{figure}
\begin{minipage}[c]{1\linewidth}
\includegraphics[width=0.49\linewidth]{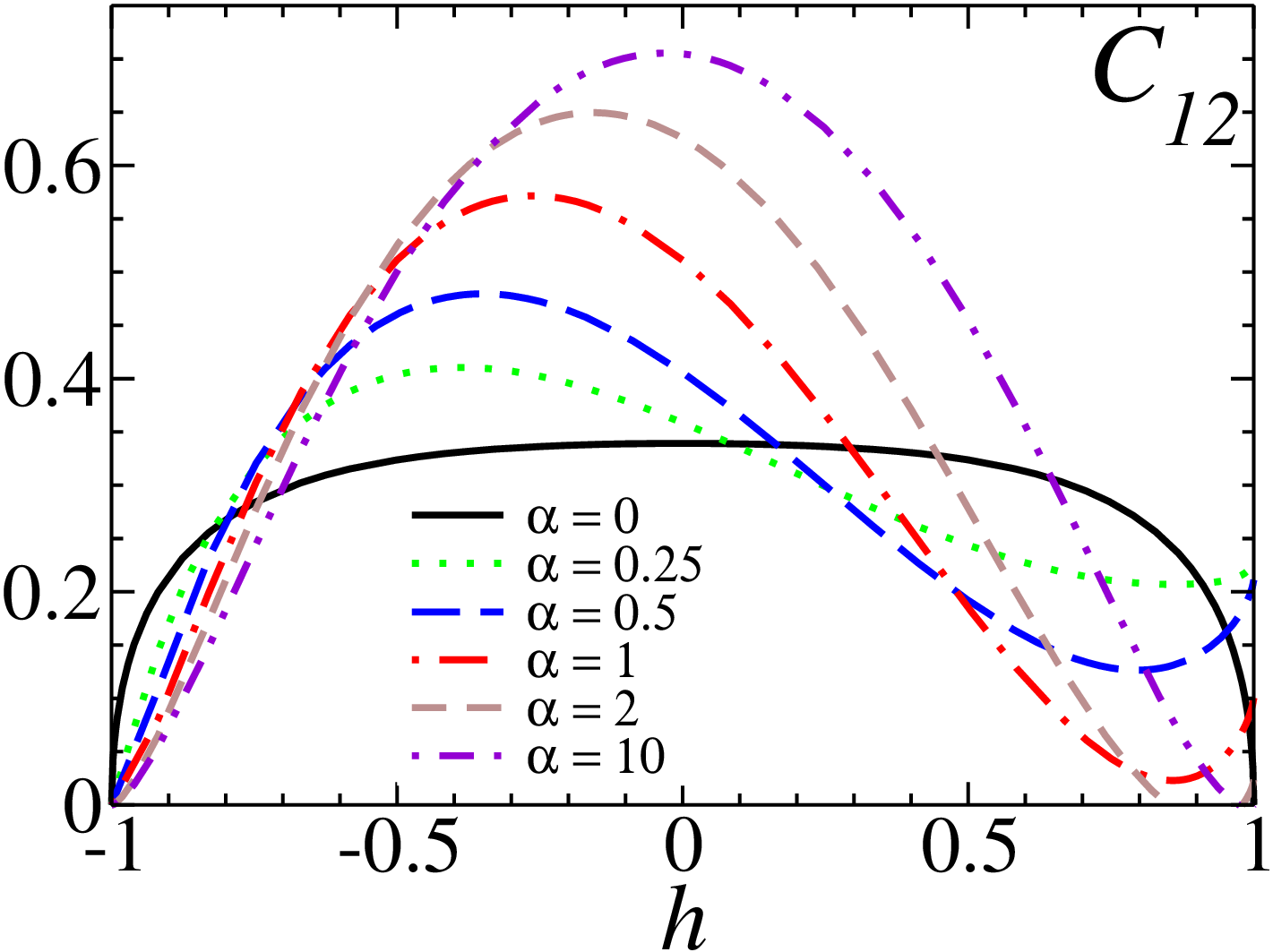}
\includegraphics[width=0.49\linewidth]{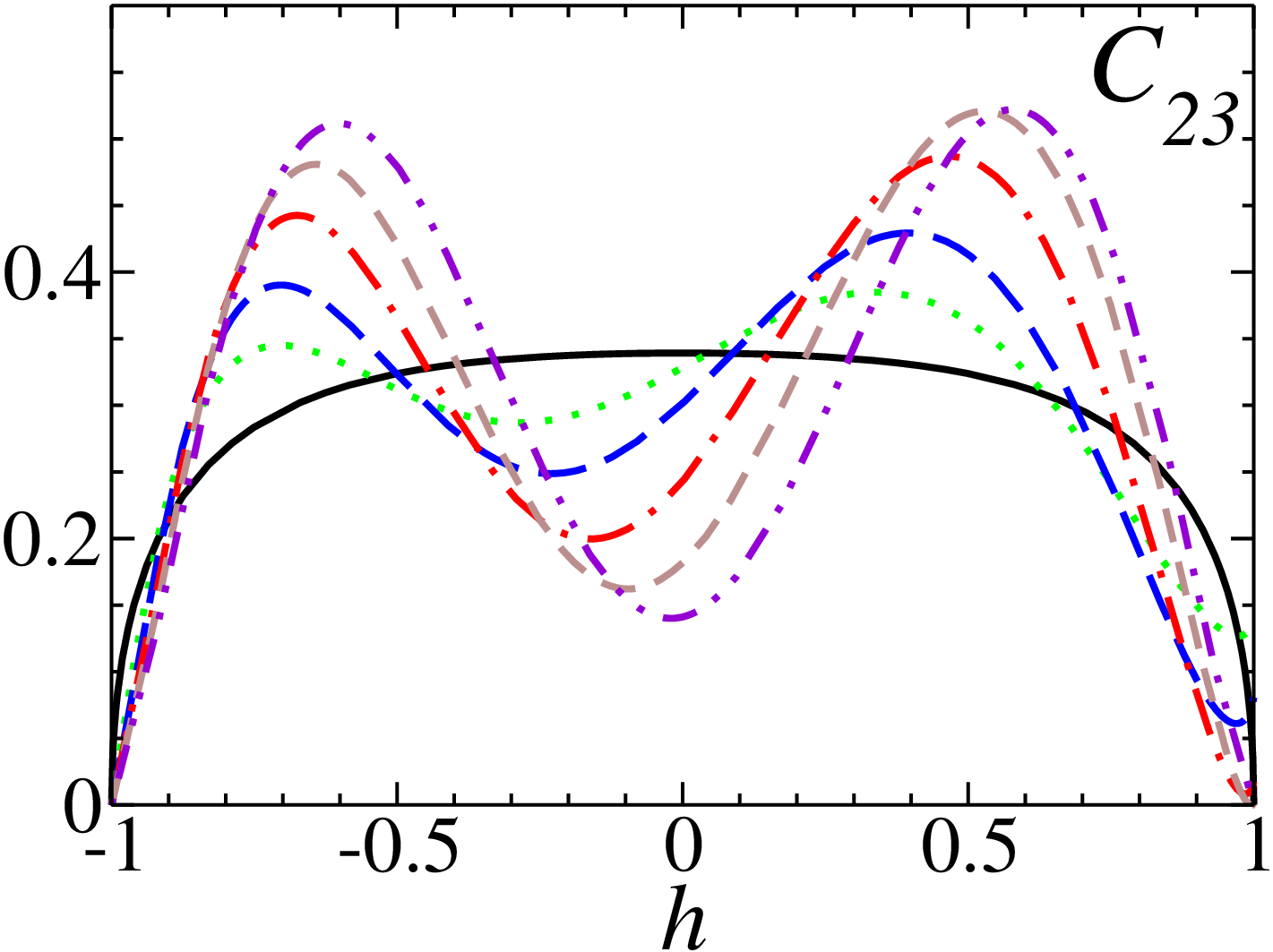}
\end{minipage}
\caption{Concurrence $C_{12}$ (left) and $C_{23}$ (right) vs $h$, for several values of $\alpha$. }
\label{f.Cs}
\end{figure}

As stated in the introduction, due to the high symmetry of its
Hamiltonian, the XX model has a markedly entangled ground state,
with next neighbor concurrences larger than those observed in less
symmetric models~\cite{AmicoEtal06}. Furthermore, the rotational
invariance forces bipartite entanglement to be of antiparallel
type~\cite{FubiniEtal06,BaroniEtal07}. This property
holds also when translational invariance is broken.
This fact, together with the emergence of the staggered structure
described above, suggests an  enhancement of the nearest neighbor
concurrences. This is shown to occur in Fig.~\ref{f.Cs}, where
$C_{12}$ and $C_{23}$ are displayed vs $h$, for different values
of $\alpha$. In particular, at zero field and for large $\alpha$,
$C_{12}$ is more than twice its value without defect, and a
similar enhancement is observed in $C_{23}$ for two different
values of $h$. In general, $C_{r,r+1}$ displays $r$ maxima, whose
position in the $[-1,1]$ field interval is increasingly symmetric,
with respect to the value $h=0$, as $\alpha$ increases. The
maximum values of $C_{r,r+1}$ are always larger than those
attained for $\alpha=0$, but the difference is smeared out with
increasing $r$ (i.e., by moving far from the impurity). The highly
non trivial behavior of $C_{r,r+1}$ with respect to $h$, $r$ and
$\alpha$, is formally due to the interference phenomenon embodied
in Eq. (\ref{dacit}). Physically it is a consequence of an
entanglement re-distribution, resulting from the constraints
imposed by the monogamy inequality~\cite{CoffmanKW00,OsborneV06}
on a locally modulated spin-structure.

In particular, for $h=0$ and independently of $\alpha$, the most
entangled spin pair is the one adjacent to the impurity and the
one-tangle $\tau_r=1-\langle{\sigma_r^z}\rangle^2$, with $r\ne 0$, does 
not vary much with $r$ and it is substantially insensitive to $\alpha$. 
The observed alternating behaviour 
of $C_{r,r+1}$, which gets a larger or smaller value with respect 
to the translational invariant case according with $r$ being 
odd or even, respectively, can thus be interpreted as a 
consequence of the monogamy inequality. Furthermore, the difference
between $C_{r,r+1}$ and $C_{r+1,r+2}$ increases with $\alpha$ and
a sort of entanglement dimerization is observed for large impurity
strength.

\begin{figure}
\begin{minipage}[c]{1\linewidth}
\includegraphics[width=0.49\linewidth]{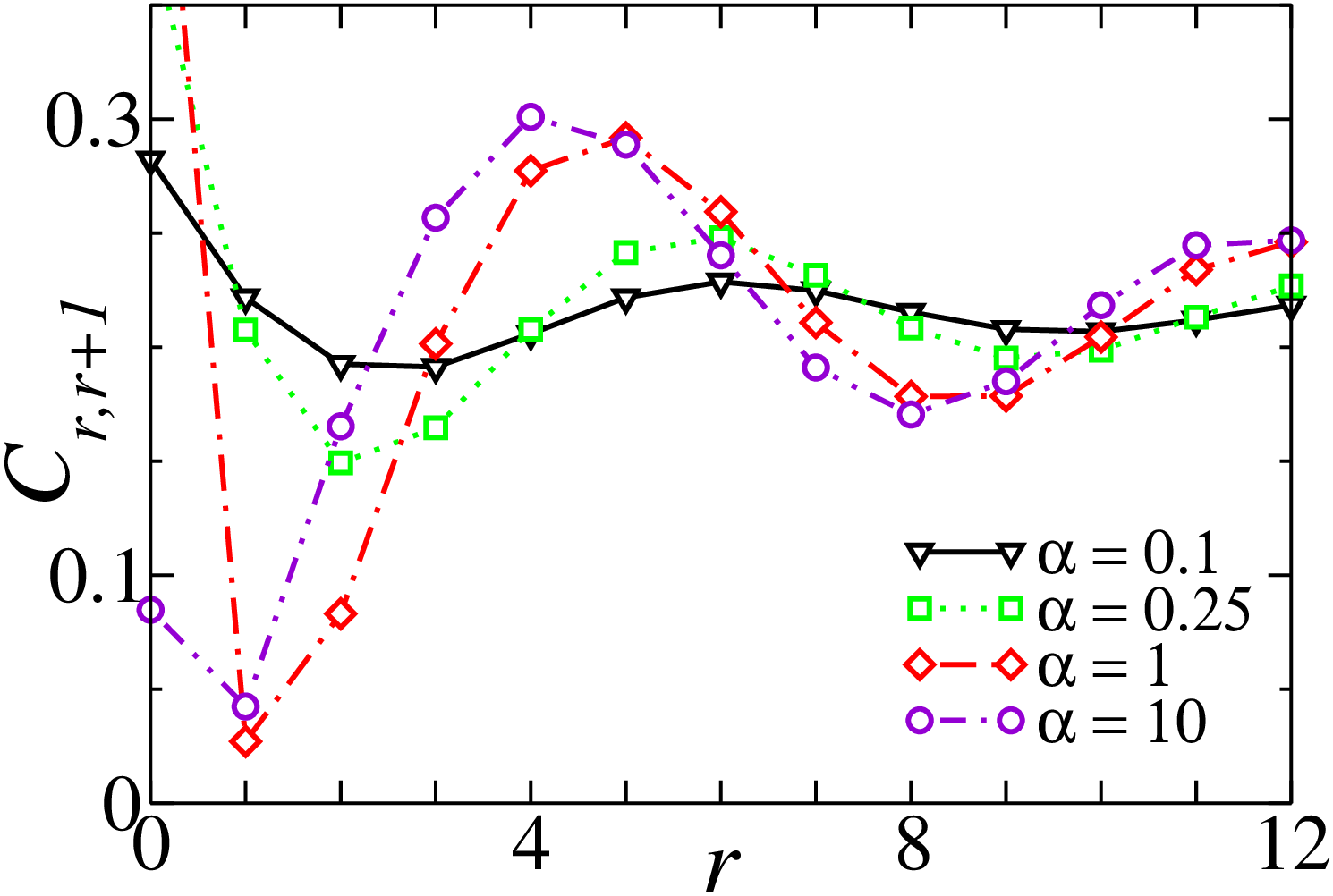}
\includegraphics[width=0.49\linewidth]{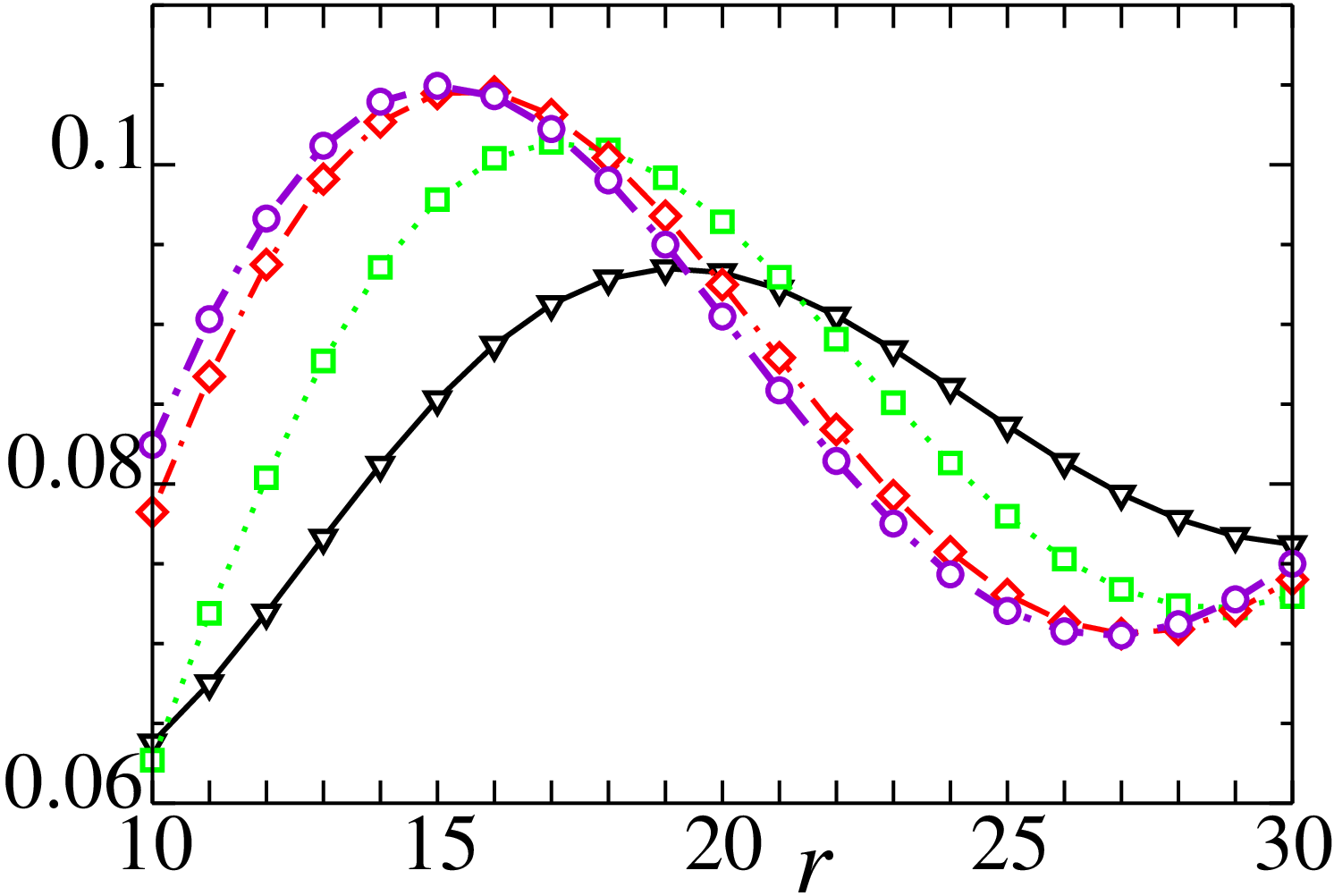}
\end{minipage}
\caption{Concurrence $C_{r,r+1}$ vs $r$ at $h=0.9$ (left), $h=0.99$ (right) for several values of $\alpha$.}
\label{conth09}
\end{figure}

For $h\neq0$, $C_{r,r+1}$ displays an increasingly complicated
dependence on $h$ and $\alpha$, which gives rise to a sharp-cut
mechanism of entanglement transfer for $h\lesssim 1$. In such case
indeed, as shown in Fig.~\ref{conth09} for $h=0.9$ and $0.99$, a
most entangled spin pair is still detected at a distance from the
impurity that increases with $h$. Hence, by varying either the
defect strength or the magnetic field, one can drive the maximum
of the concurrence along the chain, meanwhile reducing the
entanglement between spin pairs next to the defect. It is of
absolute relevance that the couplings of the spin chain with both
the uniform and the local fields are described by terms which
commute with the bare exchange XX Hamiltonian. This means that a
time variation of the control parameters $h$ and $\alpha$ induces
a fully adiabatic dynamics as non-equilibrium configurations are
never accessed.

\begin{figure}
\begin{minipage}[c]{1\linewidth}
\includegraphics[height=0.35\linewidth]{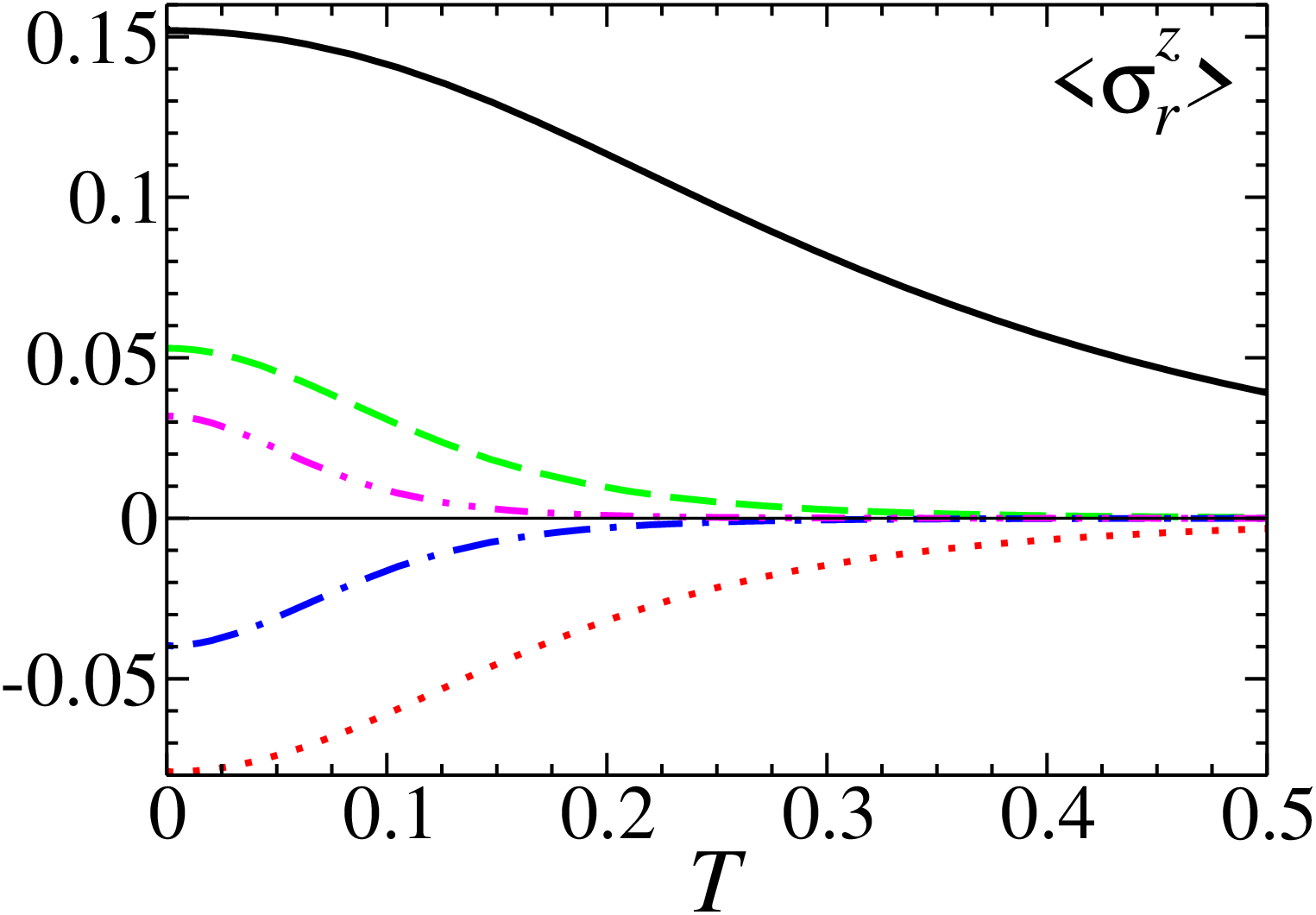}
\includegraphics[height=0.35\linewidth]{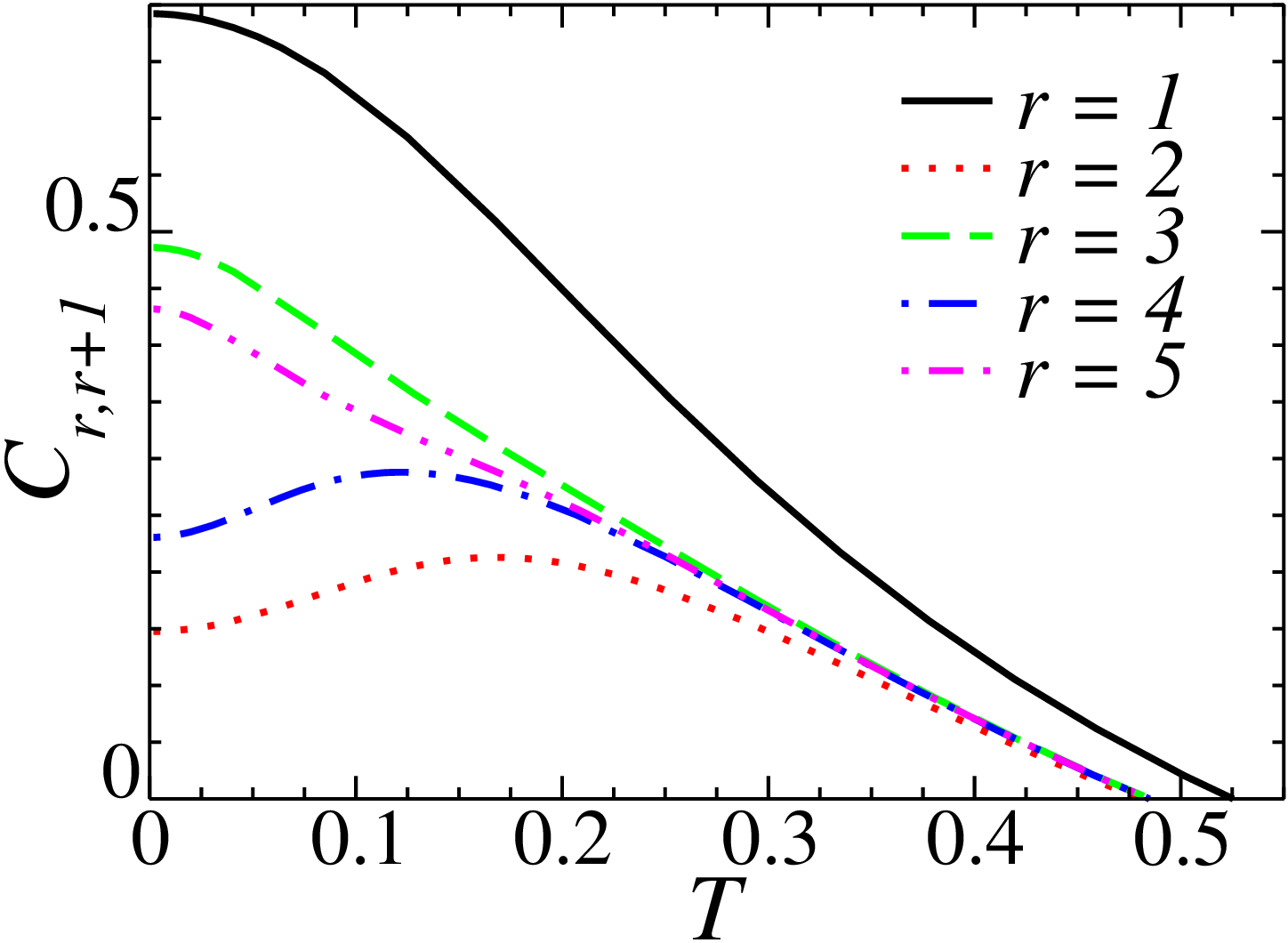}
\end{minipage}
\caption{$\langle{\sigma^z_r}\rangle$ vs $T$ for $h=0$ and $\alpha=1$ (left);
concurrence $C_{r,r+1}$ at $h=0$, $\alpha=5$ vs $T$ (right). } \label{f.Cst}
\end{figure}

When temperature $T$ is switched on, thermodynamic averages are
given by Eqs.~(\ref{e.Szz}) weighted by fermionic population
factors. As seen in Fig.~\ref{f.Cst}, the spatial distribution of
the magnetization is quite robust against temperature, essentially
because of the energy gap between the localized level and the
bottom of the band. This is also understood by noticing that the
magnetic moment at the impurity, $\langle{\sigma^z_0}\rangle$, keeps finite
at finite temperature. In fact, neglecting the contribution of the
band amplitudes, we find $\langle{\sigma^z_0}\rangle\simeq
-\frac{\alpha}{\sqrt{1+\alpha^2}}\tanh\frac{\sqrt{1+\alpha^2}}{2
T}$, where the functional dependence of the hyperbolic tangent on
$T$ accounts for the temperature robustness.

As for the concurrences, a marked dependence on $r$ survives at
$T\ne 0$ and thermal entanglement is found for selected
spin pairs, with $C_{r,r+1}$ displaying a maximum at finite
temperature for even $r$, while it decreases monotonically for $r$
odd. Although several systems which develop thermal entanglement
are known, our model provides the additional possibility to select
the spin pair for the enhancement (reduction) of entanglement.

Possible experimental studies of the above issues set in several
frameworks. First of all, quasi-one dimensional magnetic compounds
properly doped so as to introduce diluted magnetic impurities,
might be considered~\cite{DasEtal04} (to this respect, we verified that our results still hold in the case of two impurities, provided they are far enough from each other).
In this case NMR imaging of
the local magnetization near impurities could be performed, in
analogy with the case of non-magnetic
impurity~\cite{TakigawaEtal97,FujiwaraEtal98,TedoldiSH99}; the
same technique could be used to probe the spatially modulated
susceptibility~\cite{BobroffEtal97}. Moreover, muon-spin
relaxation experiments have been extensively performed to
investigate the effects of magnetic doping in high-T$_{\rm c}$
cuprates~\cite{AdachiEtal04} and might also be proper tools for
testing our findings, since the muon acts as both the impurity and
the probe for the local magnetic
arrangement~\cite{ChakhalianEtal03}. In this framework, however,
no external tuning of the parameter $\alpha$ is possible, as its
value would result from the magnetic moment of the impurity,
uniquely fixed by the dopant properties. Nonetheless, control upon
the uniform field would still be feasible. Another possible
implementation
is suggested by the recent proposal~\cite{HartmannBP07} for
realizing effective spin systems by coupled micro-cavities: in
this case, the uniform magnetic field might be tuned by varying
the frequencies of the driving lasers, while the impurity term
could result from the level structure of the micro-cavity
representing the spin sitting at the site of the impurity.
Atomic condensates in optical lattices can also be considered,
as it has been shown that they can simulate bosonic and
fermionic models which can be mapped to the XX model
in a transverse field in its quasi-long-range ordered
phase ~\cite{reticoliottici}.

This work was supported by MUR under the 2005-
2007 PRIN-COFIN National Research Projects Pro-
gram, N. 2005029421.


\begin{thebibliography}{00}
\bibitem{note1}
Alternatively, the $\alpha\, \sigma^z_0$ term may
represent an actual magnetic defect, by this meaning that the
Land\`e factor $g_{_{\rm L}}$ is not uniform along the chain (in
this latter situation $h=0$ implies $\alpha=0$).
\end{thebibliography}
\end{document}